\documentstyle[12pt] {article}
\newcommand{\beq}{\begin{equation}}
\newcommand{\eeq}{\end{equation}}
\newcommand{\beqa}{\begin{eqnarray}}
\newcommand{\eeqa}{\end{eqnarray}}
\newcommand{\ba}{\begin{array}}
\newcommand{\ea}{\end{array}}

\begin{document}

\begin{flushright}
Preprint DFPD/94/TH/57
\end{flushright}

\begin{center}
{\large \bf Charge Exchange Processes between\\
Excited Helium and Fully Stripped Ions}
\end{center}
\vspace{0.5 cm}

\begin{center}
{\bf Luca Salasnich}  \\
Dipartimento di Fisica "Galileo Galilei" dell'Universit\`a di Padova, \\
Via Marzolo 8, I 35131 Padova, Italy\\
and\\
Dipartimento di Fisica dell'Universit\`a di Firenze, \\
INFN, Sezione di Firenze, \\
Largo E. Fermi 2, I 50125 Firenze, Italy \\
\end{center}

\vskip 0.5 truecm

\begin{center}
{\bf Fabio Sattin}  \\
Istituto Gas Ionizzati del CNR, Associazione Euratom-ENEA-CNR \\
Corso Stati Uniti 4, I 35020 Padova, Italy \\
\end{center}

\vskip 0.8 truecm

\begin{center}
{\bf Abstract}
\end{center}
\vskip 0.3 truecm
\par
We made a classical trajectory Monte Carlo (CTMC) calculation of
state selective cross sections for processes between some light ions
and excited helium. The results, useful for analysis of spectroscopic
data of fusion devices, are in good agreement with
theoretical predictions of scaling laws.

\vskip 1. truecm

\begin{center}
To be published in Phys. Rev. A
\end{center}

\newpage

\par
Impurities in high temperature plasma devices are mostly found as fully
stripped ions. No spectral line is emitted by these ions, but they
may be detected through characteristic lines emitted by one--electron ions
found in excited states after a process of charge exchange (CX) with a
neutral atom:
\beq
N+X^q \to N^+ + X^{q-1}(n') \to N^+ + X^{q-1}(n) +h\nu
\eeq
where $N$ is the neutral atom, $X$ is the ion, $n$ and $n'$ are
quantum numbers, and $h\nu$ is the emitted photon.
\par
In fusion devices beams of hydrogen, helium and lithium are usually
injected into the plasma to heat it or to monitor such plasma parameters
as ion density, temperature, velocity of rotation [1]. Extraction of useful
informations from emission measurements relies on knowledge of the
effective cross section for process (1).
\par
Process (1) may take place with $N$ in the ground or in an excited
state: charge exchange involving neutrals in the ground state predominantly
populate hydrogen--like states with principal quantum number $n \leq 5$
for low--Z ions, up to oxygen [1]. The decay of these states occurs mostly
through $\Delta n=1$ transitions, corresponding to emission in the
$X$--ray or ultraviolet region [1]. The emission in the
visible region is originated from states with higher $n$
(e.g. transitions $10\to 9$ at $6070$ $\AA$ of $O^{7+}$ or $8 \to 7$ at
$5292$ $\AA$ of $C^{5+}$); in this spectral range the detection of radiation
is simpler and multichord systems, to spatially resolve the emission, are
usually working. Cross sections for electron capture into these
states from ground state neutrals are much smaller, but, on the other
hand, the contribution from excited neutrals is greatly enhanced [1] so,
even if the atoms excited by Coulomb collisions with plasma particles
constitute a small fraction of the entire beam,
they may be the major responsible for
the emission. This implies that to deduce informations from CX emission
it is necessary to have data on the processes involving excited neutrals.
\par
In this work we propose to calculate partial cross sections for process
(1) (because of their much smaller cross section double electron
processes are not here considered)
between fully stripped ions and helium in the first excited
state.
We focus on conditions frequently encountered in plasma devices : the
collision energy is varied in the range below $110$ KeV/amu and the ionic
charge varies from $3$ to $8$.
\par
A fully quantum mechanical treatment of the scattering process for such
a system would be difficult because of the large number of states
involved. The classical trajectory Monte Carlo (CTMC) method , where
the quantum system is replaced by a classical one  with all the particles
interacting through Coulomb forces and their motion as governed by
Newton's law, has proven to be a feasible and reliable tool in this
kind of calculations.
\par
The system helium$+$ion actually represents a four--body problem, but
the excited electron of helium plays a major role,
so we chose to substitute the other electron and
the nucleus altogether with a point-like particle interacting with the
other charges through a central model potential [2], which in atomic units
(a.u.) can be written:
\beq
V_{mod}(r)=-{1\over r} - {(1+0.4143 r)\over r} \exp{(-2.499 r)}.
\eeq
The first eigenvalues of the Schr\"odinger equation with this potential
are $-0.904$ a.u. and $-0.156$ a.u.. It may be proved,
following similar lines as in [3], that $V_{mod}$ provides a good
approximation of electron moment distribution, which is a critical
requisite for every CTMC computation [3,4].
\par
Initialization of the system was performed according to the procedure
developed by Reinhold and Falcon [3]. The initial distance between
nuclei was set to $10 Z$, but checks were made to larger distances,
up to $20 Z$, to ensure independence of the results from this parameter.
For each energy a number of runs ranging from
$4\cdot 10^3$ to $27\cdot 10^3$ was made to have a
reasonable statistical error ($ < 10$ per cent for the processes characterized
by the greater frequencies). We have considered ions
which are of interest in fusion devices: carbon and oxygen usually
constitute the most abundant impurities, while beryllium and boron
are often used to condition vacuum chamber inner walls.
Only distribution on the principal quantum number $n$ was considered: at
the densities and quantum numbers considered there is a complete
$l$--mixing before spontaneous decay take place [5].
\par
In Fig. 1 are plotted electron capture cross sections {\it vs} $n$ for
different energies. The position of the maximum in the
$n$--distribution of the cross section closely follows the rule:
\beq
n_{max}={Z^{3/4}\over \sqrt{-2 I}},
\eeq
where $I=-0.156$ a.u. is the ionization potential of the helium. Relation
(3) is already known to hold for collisions between ions and H or Li beams.
\par
If compared with analogue curves plotted in [6] relatively to
collisions between ions and He in the ground state, we notice that our
distributions are much broader. The spreading in $n$ is due the
increasing of the ratio $v_p/v_e$, where $v_p$ is the impact velocity of
ion, and $v_e$ is the initial orbital velocity of the electron [5].
\par
Scaling formulas with respect to charge of the ion, principal quantum number
of the neutral, and relative energy of incidence have been derived for
electron capture processes involving hydrogen. We verified
all our data lie on a single curve after the rescaling
\beq
\sigma{'}(E')={\sigma (E)\over n^4 Z},\;\;\;\;
E'={n^2 E\over \sqrt{Z}},
\eeq
(see Fig. 2). Scaling (4) has been
suggested, for instance, in [7]. Our scaling well resembles those obtained
in similar calculations with H or He in the ground state (for a comparison
see the curves plotted in [6]). In our case $n$ in eq. (4) is the effective
(noninteger) quantum number.
\par
In alternative to process (1) single ionization is also a possible result:
\beq
N+X^q \to N^+ + X^q  + e
\eeq
For ionization a similar relation to (4) has been suggested in [8] to hold,
as a result of a close-coupling calculation on ground state
and excited lithium:
\beq
\sigma{'}_{ion}(E')={\sigma_{ion} (E)\over n^4 Z^{1.3}},\;\;\;\;
E'={n^2 E\over \sqrt{Z}},
\eeq
Even though our plots cannot discriminate little differences in the exponent
of $Z$, we may state that our results are very similar to those of [8]; our
reduced cross sections are plotted in Fig. 3.
\par
In order to make sure that $V_{mod}$ may represent a good enough
potential we repeated some of the calculations by using a more sophisticated
potential: in Fig. 4 we plotted partial cross sections for the scattering
with Beryllium at $10$ KeV/amu as calculated with $V_{mod}$ and with the
model potential reported in references [9,10], the agreement being quite good.
\par
Summarizing, we have extended a well known method of calculation to
processes not previously considered, which are of importance for the
diagnostics of high temperature plasmas.
Our results are in good accordance with extrapolations from previous
calculations and empirical laws, so we may be confident on their accuracy.

\vskip 0.5 truecm
\begin{center}
{* * * * *}
\end{center}
\par
This work has been partially supported by the Ministero dell'Universit\`a
e della Ricerca Scientifica e Tecnologica (MURST). We wish to thank Dr.
M.E. Puiatti and Dr. L. Carraro for their careful reading of the manuscript.

\newpage

\parindent=0.pt

\section*{Figure Captions}
\vspace{0.6 cm}

{\bf Figure 1}: $n$--distribution for one electron capture between
helium and beryllium, boron, carbon and oxygen. Cross sections are in
$10^{-16}$ cm$^2$. Dotted curve correspond to an impact energy of $E=30$
KeV/amu, barred curve to $E=20$ KeV/amu and solid curve to $E=10$
KeV/amu.

{\bf Figure 2}: Reduced cross section $\sigma /(n^4 Z)$ for one electron
capture {\it vs} reduced energy $E n^2 / Z^{1/2}$. Cross sections are in
$10^{-16}$ cm$^2$.

{\bf Figure 3}: Reduced cross section $\sigma / (n^4 Z^{1.3})$ for single
ionization {\it vs} reduced energy $E n^2 / Z^{1/2}$.
Cross sections are in $10^{-16}$ cm$^2$.

{\bf Figure 4}: Partial cross sections $\sigma$ for single electron
capture from Beryllium at $10$ KeV/amu obtained through use of two
different model potentials. Dashed line: $V_{mod}$ of equation (2);
solid line: model potential from [9,10].
Cross sections are in $10^{-16}$ cm$^2$.
\newpage

\section*{References}
\vspace{0.6 cm}

[1] R.C. Isler: {\it Plasma Phys. Control. Fus.} {\bf 36}, 171 (1994)

[2] B.H. Bransden, A.M. Ermolaev and R. Schingal: {\it J. Phys.} B:
{\it At. Mol. Phys.} {\bf 17}, 4515 (1984)

[3] C.O. Reinhold and C.A. Falcon: {\it Phys. Rev.} A {\bf 33}, 3859 (1986)

[4] A.E. Wetmore and R.E. Olson: { \it Phys. Rev.} A {\bf 38}, 5563 (1988)

[5] R.J. Fonk, D.S. Arrow and P.K. Jaehnig: {\it Phys. Rev.} A {\bf 29},
3388 (1984)

[6] L. Meng, C.O. Reinhold and R.E. Olson: {\it Phys. Rev.} A {\bf 42}, 5286
(1990)

[7] R.E. Olson: {\it J. Phys. } B: {\it At. Mol. Phys.} {\bf 13}, 483 (1980)

[8] J. Schweinzer, D. Wutte and H.P. Winter: {\it J. Phys.} B:
{\it At. Mol. Opt. Phys.} {\bf 27}, 137 (1994)

[9] G. Peach, S.L. Willis and M.R.C. McDowell: {\it J. Phys.} B:
{\it At. Mol. Phys.} {\bf 18}, 3921 (1985)

[10] S. L. Willis, G. Peach, M.R.C. McDowell and J. Banerji:
{\it J. Phys.} B: {\it At. Mol. Phys.} {\bf 18}, 3939 (1985)

\end{document}